\documentclass[longauth]{aa} 
\usepackage{graphicx}
\usepackage{txfonts}
\usepackage{url}
\usepackage{natbib}
\usepackage{rotating}

\bibpunct{(}{)}{;}{a}{}{,}

\begin{document}
\title{Dual-frequency VLBI study of Centaurus A on sub-parsec scales}
\subtitle{The highest-resolution view of an extragalactic jet}

   \author{C. M\"uller\inst{1}\thanks{email:cornelia.mueller@sternwarte.uni-erlangen.de}
   	\and
          M. Kadler\inst{2,1,3,4}
          \and
	  R. Ojha\inst{5}
	\and
	J. Wilms\inst{1}
	\and          
         M. B\"ock\inst{1}
         \and
           P. G. Edwards\inst{6}
           \and
           C. M. Fromm\inst{7}
           \and
           H. Hase\inst{8}
           \and
           S. Horiuchi\inst{9}
           \and
           \mbox{U. Katz\inst{10}}
           \and
           J. E. J. Lovell\inst{11}
           \and
           C. Pl\"otz\inst{8}
           \and
            T. Pursimo\inst{12}
           \and
	   S. Richers\inst{1}
	   \and
           E. Ros\inst{13,7}
           \and
	   R. E. Rothschild\inst{14}
	   \and
           G. B. Taylor\inst{15}
           \and
           \mbox{S. J. Tingay\inst{16}}
           \and
           J. A. Zensus\inst{7}
          }

   \institute{ Dr. Remeis Sternwarte \& ECAP, Universit\"at Erlangen-N\"urnberg, 
             Sternwartstrasse 7, 96049 Bamberg, Germany
           \and
             Institut f\"ur Theoretische Physik und Astrophysik, Universit\"at W\"urzburg, Am Hubland, 97074 W\"urzburg, Germany
            \and
             CRESST/NASA Goddard Space Flight Center, Greenbelt, MD 20771, USA
         \and 
             Universities Space Research Association, 10211 Wincopin Circle, Suite 500 Columbia, MD 
             21044, USA
         \and
         NASA, Goddard Space Flight Center, Greenbelt, MD 20771, USA 
         \and 
              CSIRO Astronomy and Space Science, Australia Telescope National Facility, PO Box 76, Epping, NSW 1710, Australia
              \and
              Max-Planck-Institut f\"ur Radioastronomie, Auf dem H\"ugel 69, 53121 Bonn, Germany
              \and
              Federal Agency for Cartography and Geodesy, Geodetic Observatory Wettzell, Sackenrieder Str. 25, 93444 Bad K\"otzting, Germany
              \and
			CSIRO Astronomy and Space Science, CDSCC, PO Box 1035,
		Tuggeranong, ACT 2901, Australia 
              \and
              Erlangen Centre for Astroparticle Physics, Erwin-Rommel Str. 1, 91058 Erlangen, Germany
              \and
              School of Mathematics \& Physics, Private Bag 37, University of Tasmania, Hobart TAS 7001, Australia
              \and
              Nordic Optical Telescope, Apdo. de Correos 474, 38700 Santa Cruz de la Palma, Spain
              \and
              Dept. d'Astronomia i Astrof\'{\i}sica, Universitat de Val\`encia, E-46100 Burjassot, Val\`encia, Spain
              \and
              Center for Astrophysics and Space Sciences, University of California, San Diego, 9500 Gilman Drive, La Jolla, CA 92093, USA
              \and
              Department of Physics and Astronomy, University of New Mexico, Albuquerque NM, 87131, USA
              \and
              Curtin Institute of Radio Astronomy, Curtin University of Technology, Bentley, WA, 6102, Australia
             }

\abstract{Centaurus A is the closest active galactic nucleus. High resolution imaging using Very Long Baseline Interferometry (VLBI) enables us to study the spectral and kinematic behavior of the radio jet--counterjet system on sub-parsec scales, providing essential information for jet emission and formation models.}  
{Our aim is to study the structure and spectral shape of the emission from the central-parsec region of Cen\,A.}
{As a target of the Southern Hemisphere VLBI monitoring program TANAMI (Tracking Active Galactic Nuclei with Milliarcsecond Interferometry), VLBI observations of Cen\,A are made regularly at 8.4 and 22.3\,GHz with the Australian Long Baseline Array (LBA) and associated telescopes in Antarctica, Chile, and South Africa.}
{The first dual-frequency images of this source are presented along with the resulting spectral index map. An angular resolution of 0.4\,mas $\times$ 0.7\,mas is achieved at 8.4\,GHz, corresponding to a linear scale of less than 0.013\,pc. Hence, we obtain the highest resolution VLBI image of Cen\,A, comparable to previous space-VLBI observations. By combining with the 22.3\,GHz image, which has been taken without contributing transoceanic baselines at somewhat lower resolution, we present the corresponding dual-frequency spectral index distribution along the sub-parsec scale jet revealing the putative emission regions for recently detected $\gamma$-rays from the core region by \textit{Fermi}/LAT.}
{We resolve the innermost structure of the milliarcsecond scale jet and counterjet system of Cen\,A into discrete components. The simultaneous observations at two frequencies provide the highest resolved spectral index map of an AGN jet allowing us to identify multiple possible sites as the origin of the high energy emission.} 

  \authorrunning{C. M\"uller et al.}
  \titlerunning{Dual-frequency VLBI imaging of Centaurus A at sub-parsec scales}
  \maketitle
%

\section{Introduction\label{sec:intro}}

At a distance of  $3.8 \pm 0.1$\,Mpc \citep{Harris2010}, the giant elliptical galaxy Centaurus A (PKS\,1322$-$428, NGC\,5128) is the closest active galactic nucleus (AGN). Cen\,A is usually classified as a Fanaroff-Riley type I radio galaxy \citep{FanaroffRiley1974} and it exhibits double-sided kiloparsec scale jets which end in giant radio lobes \citep{Clarke1992}. These jets are produced in the vicinity of a supermassive black hole with a mass of \mbox{$M=5.5\pm3.0\times 10^7 \mathrm{M}_{\sun}$} \citep{Israel1998, Neumayer2010}.  
Due to the proximity of Cen\,A, an angular distance of one milliarcsecond (mas) corresponds to a linear size of just $\sim $$0.018$\,pc. 
This makes it an exceptionally good laboratory for studying the innermost regions of AGN, for studying jet formation and collimation as well as testing different jet emission models. Cen\,A's parsec scale structure has been 
well studied at radio frequencies using Southern-Hemisphere Very Long Baseline Interferometry (VLBI), revealing a bright jet and a faint 
counter-jet at a viewing angle of 50--80 degrees to the line of sight, with speeds ranging from  $v=0.1c$ to $0.45c$ measured on different scales \citep{Tingay1998, Tingay2001a, Hardcastle2003}. On subparsec scales, the radio jet-counterjet system was resolved with the VLBI Space Observatory Program  (VSOP) by \cite{Horiuchi2006}.

Cen\,A is detected across the spectrum. 
\textit{CGRO}/EGRET found the AGN to be a likely source of $\gamma$-rays \citep{Hartman1999} and recently, the Large Area Telescope \citep[LAT;][]{Atwood2009} detector aboard the \textit{Fermi Gamma-ray Space Telescope} (hereafter \textit{Fermi}/LAT) detected strong $\gamma$-ray emission 
from both the central (nuclear) region \citep{Abdocenacore2010} as well as the giant radio lobes \citep{Abdocenascience2010}, while TeV emission was detected by H.E.S.S. \citep{Aharonian2009}.  

The Tracking Active Galactic Nuclei with Austral Millli\-arcsecond Interferometry program \citep[TANAMI;][]{Ojha2010}\footnote{\url{http://pulsar.sternwarte.uni-erlangen.de/tanami/}} has been monitoring Cen\,A at two radio frequencies 
about two times a year since 2007 Nov. 
We present results from the TANAMI experiments in 2008 Nov (Sect.~\ref{sec:obs}). These included for the first time the TIGO and O'Higgins antennas (at 8.4\,GHz), as well as full participation of the Tidbinbilla 70\,m antenna (at 8.4 and 22.3\,GHz) throughout the experiments, resulting in both the highest resolution images of Cen\,A ever made
and the highest fidelity spectral index map of its mas jet (Sect.~\ref{sec:results}). In Sect.~\ref{sec:discuss}, we discuss the structure of the inner parsec of Cen\,A and constraints on the location of possible $\gamma$-ray emission regions.

\begin{figure}[tb]
\centering
\includegraphics[width=0.95\columnwidth]{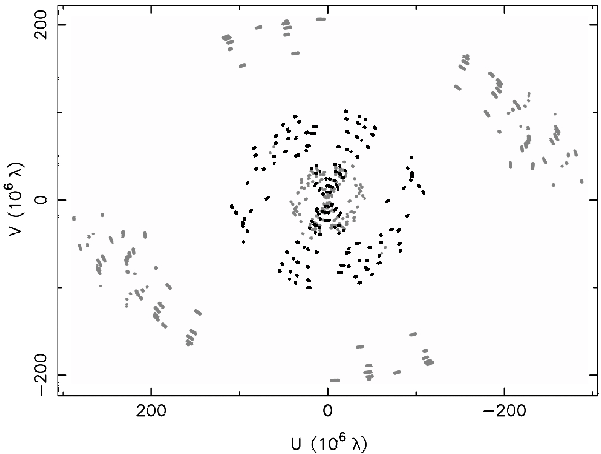}
\caption{($u$,$v$)-coverage at 8.4\,GHz (grey) and at 22.3\,GHz (black) for Cen\,A (see Table~\ref{table:2}). Note that the transoceanic antennas O'Higgins and TIGO do not support observations at 22.3\,GHz.}\label{fig:uv}
\end{figure}
\begin{table*}[htb]
\caption{Image parameters} 
\label{table:2} 
{\centering
\begin{tabular}{l c c c c c c c c c} 
\hline\hline 
Weighting & Frequency & Epoch & RMS & $S_{\mathrm{peak}}$ &  $T^\mathrm{b}_{\mathrm{B, peak}}$ & $S_{\mathrm{total}}$ & $b_{\mathrm{maj}}$ &  $b_{\mathrm{min}}$ & P.A. \\
& $[\mathrm{GHz}]$	& yyyy-mm-dd& [$\mathrm{mJy\,beam^{-1}}$] &  [$\mathrm{Jy\,beam^{-1}}$] & [$10^{10} \mathrm{K}$]& [$\mathrm{Jy}$] & [$\mathrm{mas}$] & [$\mathrm{mas}$] & [$^\circ$] \\
\hline 
{\textbf {natural}} & \phantom{0}8.4  & 2008-11-27 & $0.39\pm0.02$ &0.74 &  2.20	& $3.9\pm0.1$ &0.98	&0.59   & $\phantom{-}31$	 \\
& 22.3 & 2008-11-29 & $0.47\pm0.03$ &1.77 &  0.17	& $3.4\pm0.1$ &2.01	&1.27   & $\phantom{-}88$	 \\
\hline 
{\textbf {uniform}} & \phantom{0}8.4  & 2008-11-27 & $0.45\pm0.02$ &0.48	& 2.84 & $3.2\pm0.1$ &0.68	&0.43   & $\phantom{-}33$ 	\\     
& 22.3 & 2008-11-29 & $1.20\pm0.10$   &1.46 & 0.19 & $3.3\pm0.1$ &1.55	&1.21   & $-75$ 	 \\  
\hline 
\end{tabular}\\}
\footnotesize{Columns show: (1) weighting scheme; (2) observing frequency; (3) observation epoch; (4) RMS noise in the {\sc clean} images; (5) peak flux density; (6) brightness temperature derived from $T_\mathrm{B}=1.22\cdot10^{12}(S_\nu/\mathrm{[Jy]})(\nu/\mathrm {[Hz]})^{-2}(\theta/\mathrm{[mas]})^{-2}$ using the restoring beam \citep[][]{Condon1982}; (7) total integrated flux density; (8,9,10) major axis, minor axis, and position angle of the restoring beam}
\end{table*}

%
\section{Observations and Data Reduction \label{sec:obs}}

TANAMI observations at 8.4\,GHz (2008 Nov 27)
and 22.3\,GHz (2008 Nov 29) were made quasi-simultaneously with the Australian Long Baseline Array 
(LBA)
and associated telescopes, the 9\,m German Antarctic Receiving Station (GARS) in O'Higgins, Antarctica, the 6\,m Transportable Integrated Geodetic Observatory (TIGO) in Chile, both of which are operated by the Bundesamt f\"{u}r Kartographie und Geod\"{a}sie (BKG), Germany, and the 70\,m DSS\,43 antenna of the NASA Deep Space Network in Tidbinbilla, Australia. The LBA consists of five telescopes, Parkes (64\,m), ATCA (5$\times$22\,m),  Mopra (22\,m) all in New South Wales, Hobart (26\,m) in Tasmania, and Ceduna (30\,m) in South Australia. 
Figure~\ref{fig:uv} shows the resulting ($u$,$v$)-coverage for Cen\,A at both frequencies. Note that the resolution is {\sl lower} at the {\sl higher} frequency as GARS and TIGO cannot observe at 22.3\,GHz.
Data were recorded on LBA Disk Recorders and correlated on the DiFX software correlator \citep{Deller2007} at Curtin University in Perth, Western Australia 
and calibrated and imaged as described in \citet{Ojha2010}.

%
\section{Results \label{sec:results}}
\begin{figure*}[tb] 
\hfill\includegraphics[width=16.7cm]{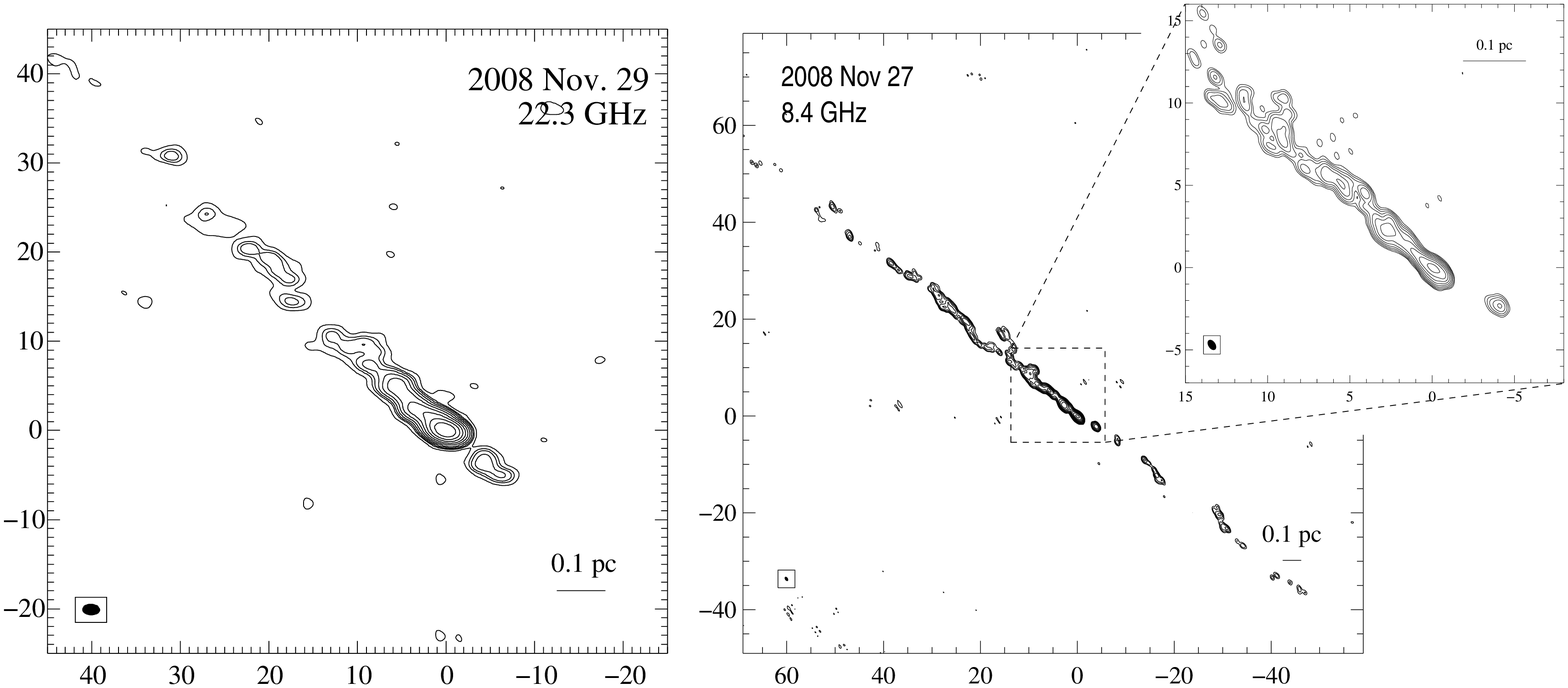}\hspace{0.5cm}
	\caption{22.3\,GHz and 8.4\,GHz images of Cen\,A from 2008 November 27 and 29, using natural weighting. The inset shows the inner region of Cen\,A at 8.4\,GHz using uniform weighting. The lowest contours denote the 3$\sigma$-noise-level (see Table~\ref{table:2}). All axes are in units of mas relative to the phase center. The ellipse in the lower left corner of each panel indicates the restoring beam.}
            \label{fig:X}
    \end{figure*}

Figure~\ref{fig:X} shows images of Cen\,A at 8.4\,GHz and 22.3\,GHz. Image parameters and observation characteristics are listed in Table~\ref{table:2}. Using uniform weighting at 8.4 GHz, an angular resolution of 0.7 $\times$ 0.4 \,mas at a position angle (P.A.) of $\sim$$33^\circ$) is achieved corresponding to linear scales of less than 3500\,AU. 

The most important constraint on the fidelity of the 8.4\,GHz image is the lack of intermediate baselines between the LBA and the two transoceanic antennas, evident from ``holes'' in the ($u$,$v$)-plane \citep[Fig.~\ref{fig:uv}; see, e.g.,][for a discussion of possible effects due to uneven coverage of the ($u$,$v$)-plane]{Lister2000}. To test the reliability of the imaging, we continuously checked whether our clean model describes the uncalibrated visibility data consistently. In addition, we gauged the robustness of features in given regions of the map by iteratively testing the effect of self-calibrating the data without including these features in our \textsc{clean} models. 
We estimate the calibration uncertainties to be on the order of 15\,\% at both frequencies. On-source errors due to the uneven ($u$,$v$)-coverage were measured following the methods described by  \citet{Tingay2001b} and found to be smaller than 15\,\% along the inner $\sim$$15$\,mas of the jet. Beyond this, the on-source errors gradually grow and can exceed 100\,\% towards the counterjet at 8.4\,GHz.   
We also compared our final images with previous and subsequent, somewhat lower resolution, TANAMI observations \citep{Mueller2010} by identifying the same main bright features  to achieve a self-consistent multi-epoch kinematics model which will be  analyzed in a subsequent paper.

At both frequencies, our images show a well collimated jet with an opening angle $\theta\lesssim12^\circ$ at a mean P.A. of $\theta\sim$$50^\circ$ and a fainter counterjet ($\theta\sim$$-130^\circ$) with an emission gap in between. At 22.3\,GHz, the counterjet is considerably weaker: 
for the 8.4\,GHz counterjet feature around position $\textrm{R.A.}=-18$, $\textrm{Dec}=-14$, we put a limit of the spectral index $\alpha < -2.5$ (defined via \mbox{$F_\nu \propto \nu^{+\alpha}$}). 
Both the P.A. 
and the opening angle are remarkably similar to the corresponding values of $P.A._\mathrm{arcmin}=50^\circ\mathrm{-}60^\circ$ and  $\theta_\mathrm{arcmin}\sim$$12^\circ$ seen on scales of $\lesssim 4$\,arcmin ($4$\,kpc) from the central engine \citep{Clarke1992}. 
The feature with the highest peak-flux density (see Table~\ref{table:2}) has an inverted spectrum (see below) and therefore shares the characteristic properties of the compact feature at the upstream-end of most blazar jets, typically called ``the core'' of the jet. Note that in the case of Cen\,A our linear resolution is orders of magnitude better than in most blazar-jet images and consequently most of the structure seen here within the central parsec would be contained in ``the core'' if Cen\,A was observed at a smaller angle to the line of sight or at larger distance. 
The significant emission features within the sub-parsec scale jet seen in the 2007 November 8.4\,GHz TANAMI image \citep{Mueller2010, Ojha2010} are in good agreement with those in this image observed one year later. 
A possible widening and subsequent narrowing of the jet appears at 
$\sim$25\,mas ($\approx 0.45$\,pc), and also in the earlier lower resolution image, as well as in the 22.3\,GHz image (see Fig.~\ref{fig:X}). The peak-flux densities at 8.4\,GHz are comparable, 0.6\,Jy in 2007 November and 0.7\,Jy in 2008 November, indicating only a moderate radio flux variability. The 2008 November image reveals a small counterjet displacement from the jet axis. The overall jet structure is also similar to the earlier 5\,GHz VSOP image \citep{Horiuchi2006}. This similarity suggests that at least some characteristic structures are moving slowly compared to the jet speed  measured in earlier works \citep[$0.1c$--$0.45c$;][]{Tingay1998, Tingay2001a, Hardcastle2003}. 

The spectral distribution along the jet constrains the physical properties of the radio emission. The synthesis imaging self-calibration procedure leads to the loss of the absolute source position. Therefore, the frequency-dependent VLBI shift of self-absorbed, optically thick emission features \citep[e.g.,][]{Lobanov1998} has to be taken into account to assemble a spectral index map. We determined the offset by aligning the positions of optically thin Gaussian model components, which are not affected by the shift, to be $\Delta\alpha_\mathrm{rel}=-0.25$\,mas and $\Delta\delta_\mathrm{rel}=-0.2$\,mas.
	\begin{figure}[htb] 
	\resizebox{0.87\hsize}{!}{\includegraphics{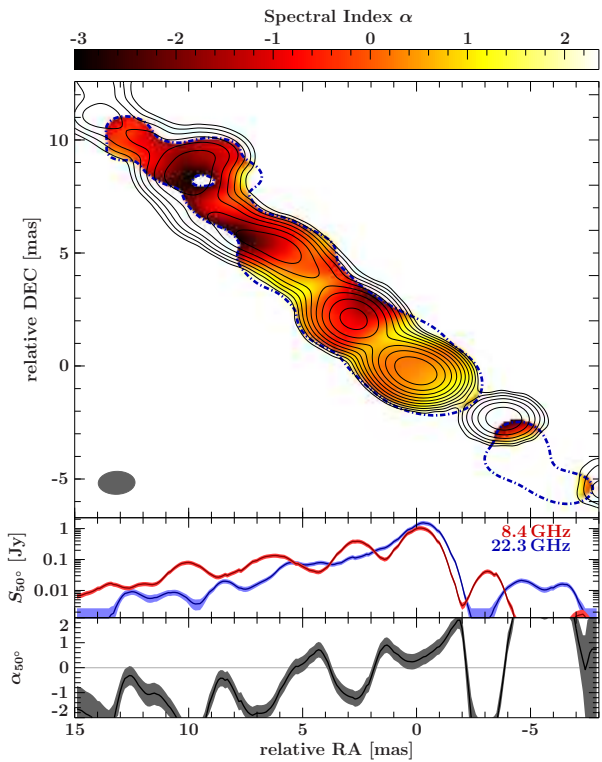}}
	\caption{Top: Spectral index map derived from the flux densities at 8.4\,GHz and 22.3\,GHz. The flux density distribution at 8.4\,GHz is shown in solid black contours. The dashed blue contour shows the 3$\sigma$-noise level at 22.3\,GHz. Both maps have been restored with a beam of $1.61 \times 1.02$\,mas ($\theta=88^\circ$) represented by the gray ellipse in the lower left corner. Middle: Flux density distribution at 8\,GHz (red) and 22.3\,GHz (blue) for a narrow strip of 0.1\,mas along the jet at a P.A.=$50^\circ$ corrected for the shift of $\Delta\alpha = - 0.25$\,mas and $\Delta\delta = - 0.2$\,mas. Bottom: Spectral index given by the flux densities at P.A.=$50^\circ$. Uncertainties corresponding to the absolute calibration uncertainties and on-source errors of $\sim$$15$\,\% and considering the image rms at both frequencies are shown as shaded regions around the best-fit distributions (solid lines). 
	\label{fig:spixmap}}
	\end{figure}

\section{Discussion and Conclusions\label{sec:discuss}}

Recently, \textit{Fermi}/LAT has detected  $\gamma$-ray emission from the core region of Cen\,A  \citep{Abdocenacore2010} but 
the LAT positional accuracy does not allow to locate the $\gamma$-ray emission  to an accuracy of better than $\sim$$0\fdg1$.
Our simultaneous dual-frequency VLBI data allow us to study spectral changes at sub-parsec scale and therefore to constrain the possible origin of the $\gamma$-ray emission from the core region of Cen\,A. To do this,
we fit Gaussian model components to the visibility data (see Table~\ref{table:3}). 
We can clearly resolve the 7\,mas large core region imaged by \citet{Horiuchi2006} into two separate jet features.  
The brighter feature at the base of the jet, with an angular extent of $\sim$3\,mas, can be described with three Gaussian model components at the higher-resolution 8.4\,GHz image while two components are sufficient to represent the lower-resolution 22.3\,GHz core region. We identify this region as the innermost visible part of the jet at this frequency (more central parts being self-absorbed). The second bright emission feature at $\sim$$3.5$\,mas downstream at both frequencies was also detected by TANAMI in 2007 November \citep{Ojha2010} and might correspond to the stationary component detected by \cite{Tingay2001a} at $\sim$$4$\,mas.  Co-spatial with the jet-widening region at $\sim$23 mas at 8.4\,GHz, the 22.3\,GHz image suggests a possible emission component, which might be interpreted as an absorbing feature. However, the limited $(u,v)$-coverage does not allow us to fully constrain the exact spatial distribution of this emission feature and consequently spatially resolved spectroscopic analysis has to be treated with caution.  

Figure~\ref{fig:spixmap} represents the  first sub-parsec scale dual-frequency image of Cen\,A. The core region has an inverted spectrum (spectral index $0 < \alpha < 2$) which changes from flat to steep downstream. Remarkably, the spectrum remains flat for the inner few milliarcseconds of the jet, indicating multiple optically thick emission regions in the core region although a gradual steepening behaviour can be seen. These optically thick regions can be \emph{identified as multiple possible production sites of the high energy photons} because they all show high brightness temperatures, compact, structures, and flat radio spectra. This should be considered in the discussion of single-zone versus multi-zone emission models of the broadband spectral energy distribution of Cen\,A and other extragalactic jets. Recent studies of Cen\,A \citep{Abdocenacore2010, Roustazadeh2011} modeling the full radio to TeV range of the broadband spectrum have used TANAMI spectral data from the inner $0.9\textrm{\,mas} \times 0.29\textrm{\,mas}$\footnote{Note the slightly different model component values as compared to the values reported in \cite{Abdocenacore2010}, which reflect self-calibration and model-fitting uncertainties.}.  

As Fig.~\ref{fig:spixmap} shows, the peak of the innermost counterjet feature at 22.3\,GHz is offset by about 1\,mas to the south-west with respect to the 8\,GHz image. This offset could be interpreted as a sign of free-free absorption in an obscuring torus as observed in other two-sided extragalactic jets \citep[e.g.,][]{Kadler2004}. In this case, however, it is puzzling that the  well-defined 8\,GHz component at $\Delta\alpha=-4$\,mas, $\Delta\delta=-2$\,mas seems to have no obvious counterpart at 22.3\,GHz. Free-free absorption has been observed before in the central region of Cen\,A \citep{Tingay2001b}. In fact, this may also affect the innermost part of the eastern jet and could be an alternative explanation for the observed flat spectral index distribution.

In contrast to M87 \citep{Junor1999}, no rapid broadening of the jet opening angle is seen, although the linear resolutions are comparable. 
The jet of Cen\,A is therefore already well collimated on scales of $0.1\textrm{\,pc}=10^{15}\textrm{\,m}$. Beyond this distance, the opening angle is $\sim$$12^\circ$, about the same value as measured on kpc-scales \citep{Clarke1992}. However, the mass of the central black hole of M87 is two orders of magnitude higher than that of Cen\,A so we are accessing a very different part of the jet as measured in Schwarzschild radii. 

We plan to determine the evolution of spectral index and the proper motion using observations from multiple TANAMI epochs. Combining this with $\gamma$-ray monitoring, this approach will allow us to further constrain the possible $\gamma$-ray emission regions. 
Future ultra-high resolution multi-frequency observations with a global VLBI array
(including the southernmost VLBA antennas) will let us measure the spectral turnover frequency and probe the presence of a dense free-free absorber. In combination with kinematic data from the TANAMI observations, we will be able to estimate the magnetic field and relativistic electron energy density, key parameters of jet broadband emission models. 

\begin{table}[htb]
\caption{Modelfit parameters for core region (natural weighting)} 
\label{table:3} 
{\centering
\begin{tabular}{@{}c c c c c c c@{}} 
\hline\hline 
Flux density & Radius & $\theta$ & $b_\mathrm{maj}$ &  $b_\mathrm{min}$ & P.A. & $T_{\mathrm{B}}$\\
$[\mathrm{Jy}]$ &  [mas] & [$^\circ$] & [$\mathrm{mas}$] & [$\mathrm{mas}$] & [$^\circ$] & [$10^{10} \mathrm{K}$]\\
\hline 
{\textbf {8.4\,GHz}}\\
0.493 &     0.61 & $-$121.5 &  0.32  &   0.32 &    \phantom{$-$}60.8 &	 \phantom{0}8.23 \\
0.772 &     0.05 & \phantom{$-$}105.9 &  0.30  &   0.30 &    \phantom{$-$}65.5 &	 14.83\\
0.335 &     0.71 & \phantom{$-$1}64.6 &  0.18  &   0.18 &    \phantom{$-$}36.5 &  17.87	\\ 
\hline 
{\textbf{22.3\,GHz}}\\
1.620 &    0.57&     $-$131.9 &     0.77 &     0.17 &     \phantom{$-$}60.2 & 3.04    \\
0.809 &    0.57&     \phantom{$-$1}60.0 &     0.59 &     0.87 &    $-$80.3 & 0.39\\
\hline 
\end{tabular}}
\end{table}
\vspace*{-\baselineskip}
\begin{acknowledgements}

We acknowledge partial funding from the Bun\-des\-mini\-sterium f\"ur Wirtschaft und Technologie under Deutsches Zentrum f\"ur Luft- und Raumfahrt grant 50 OR 0808 and from the European Commission under contract number ITN 215212 ``Black Hole Universe''.
This research was funded in part by NASA through Fermi Guest Investigator grant NNH09ZDA001N. This research was supported by an appointment to the NASA Postdoctoral Program at the Goddard Space Flight Center, administered by Oak Ridge Associated Universities through a contract with NASA. E.R. acknowledges partial support by the Spanish MICINN through grant AYA2009-13036-C02-02.
The Long Baseline Array is part of the Australia Telescope National Facility which is funded by the Commonwealth of Australia for operation as a National Facility managed by CSIRO.
We thank the referee, D.\,L.\,Jauncey, for insightful comments which helped us to improve the manuscript. 
\end{acknowledgements}


\begin{thebibliography}{}

\bibitem[\protect\astroncite{{Abdo} et~al.}{2010a}]{Abdocenacore2010}
{Abdo} A.A., {Ackermann} M., {Ajello} M., et~al., 2010a, ApJ 719, 1433

\bibitem[\protect\astroncite{{Abdo} et~al.}{2010b}]{Abdocenascience2010}
{Abdo} A.A., {Ackermann} M., {Ajello} M., et~al., 2010b, Sci 328, 725

\bibitem[\protect\astroncite{{Aharonian} et~al.}{2009}]{Aharonian2009}
{Aharonian} F., {Akhperjanian} A.G., {Anton} G., et~al., 2009, ApJ 695, L40

\bibitem[\protect\astroncite{{Atwood} et~al.}{2009}]{Atwood2009}
{Atwood} W.B., {Abdo} A.A., {Ackermann} M., et~al., 2009, ApJ 697, 1071

\bibitem[\protect\astroncite{{Clarke} et~al.}{1992}]{Clarke1992}
{Clarke} D.A., {Burns} J.O., {Norman} M.L.,  1992, ApJ 395, 444

\bibitem[\protect\astroncite{{Condon} \& {Mitchell}}{1982}]{Condon1982}
{Condon} J.J., {Mitchell} K.J.,  1982, AJ 87, 1429

\bibitem[\protect\astroncite{{Deller} et~al.}{2007}]{Deller2007}
{Deller} A.T., {Tingay} S.J., {Bailes} M., {West} C.,  2007, PASP 119, 318

\bibitem[\protect\astroncite{{Fanaroff} \& {Riley}}{1974}]{FanaroffRiley1974}
{Fanaroff} B.L., {Riley} J.M.,  1974, MNRAS 167, 31p

\bibitem[\protect\astroncite{{Hardcastle} et~al.}{2003}]{Hardcastle2003}
{Hardcastle} M.J., {Worrall} D.M., {Kraft} R.P., et~al., 2003, ApJ 593, 169

\bibitem[\protect\astroncite{{Harris} et~al.}{2010}]{Harris2010}
{Harris} G.L.H., {Rejkuba} M., {Harris} W.E.,  2010, Proc. Astro. Soc. Aus. 27,
  457

\bibitem[\protect\astroncite{{Hartman} et~al.}{1999}]{Hartman1999}
{Hartman} R.C., {Bertsch} D.L., {Bloom} S.D., et~al., 1999, ApJS 123, 79

\bibitem[\protect\astroncite{{Horiuchi} et~al.}{2006}]{Horiuchi2006}
{Horiuchi} S., {Meier} D.L., {Preston} R.A., {Tingay} S.J.,  2006, PASJ 58, 211

\bibitem[\protect\astroncite{{Israel}}{1998}]{Israel1998}
{Israel} F.P.,  1998, AARv 8, 237

\bibitem[\protect\astroncite{{Junor} et~al.}{1999}]{Junor1999}
{Junor} W., {Biretta} J.A., {Livio} M.,  1999, Nat 401, 891

\bibitem[\protect\astroncite{{Kadler} et~al.}{2004}]{Kadler2004}
{Kadler} M., {Ros} E., {Lobanov} A.P., et~al., 2004, A\&A 426, 481

\bibitem[\protect\astroncite{{Lister} et~al.}{2000}]{Lister2000}
{Lister} M.L., {Piner} B.G., {Tingay} S.J.,  2000,
\newblock In: {H.~Hirabayashi, P.~G.~Edwards, \& D.~W.~Murphy} (ed.) Astrophys.
  Phenomena Revealed by Space VLBI, Sagimahara (Japan)., p.189

\bibitem[\protect\astroncite{{Lobanov}}{1998}]{Lobanov1998}
{Lobanov} A.P.,  1998, A\&AS 132, 261

\bibitem[\protect\astroncite{{M{\"u}ller} et~al.}{2010}]{Mueller2010}
{M{\"u}ller} C., {Kadler} M., {Ojha} R., et~al., 2010,
\newblock In: {Savolainen, T., Ros, E., Porcas, R.W. \& Zensus, J.A.} (ed.)
  Proc. "Fermi meets Jansky -- AGN in Radio and Gamma-Rays", Bonn (Germany):
  MPIfR., p.229

\bibitem[\protect\astroncite{Neumayer}{2010}]{Neumayer2010}
Neumayer N.,  2010, Proc. Astro. Soc. Aus. 27, 449

\bibitem[\protect\astroncite{{Ojha} et~al.}{2010}]{Ojha2010}
{Ojha} R., {Kadler} M., {B{\"o}ck} M., et~al., 2010, A\&A 519, A45

\bibitem[\protect\astroncite{{Roustazadeh} \&
  {B{\"o}ttcher}}{2011}]{Roustazadeh2011}
{Roustazadeh} P., {B{\"o}ttcher} M.,  2011, ApJ 728, 134

\bibitem[\protect\astroncite{{Tingay} et~al.}{1998}]{Tingay1998}
{Tingay} S.J., {Jauncey} D.L., {Reynolds} J.E., et~al., 1998, \aj 115, 960

\bibitem[\protect\astroncite{{Tingay} \& {Murphy}}{2001}]{Tingay2001b}
{Tingay} S.J., {Murphy} D.W.,  2001, ApJ 546, 210

\bibitem[\protect\astroncite{{Tingay} et~al.}{2001}]{Tingay2001a}
{Tingay} S.J., {Preston} R.A., {Jauncey} D.L.,  2001, AJ 122, 1697

\end{thebibliography}

\end{document}